\documentclass[twoside]{article}

\evensidemargin\oddsidemargin
\advance \baselineskip 0pt plus 0.1pt minus 0.1pt
\input epsf
\usepackage[T2A]{fontenc}

\usepackage{latexsym}
\usepackage{amsbsy}
\usepackage{amsfonts}
\usepackage{amssymb}

\usepackage[mathcal]{eucal}

\date{}

\begin{document}
\def\Z{\hbox{{\sf Z}\kern-0.4em {\sf Z}}}

\title{Toroid Current States in Doped Mott Insulators}

\author{Dmitry I. Iudin${ }^{1}$,
Alexander P. Protogenov${ }^{1,2}$\footnote{e-mail: alprot@appl.sci-nnov.ru}  \\ \\
{\fontsize{10pt}{12pt}\selectfont
${}^{1}${\em Institute of Applied Physics of the RAS, Nizhny Novgorod 603950, Russia
}\/}\\
{\fontsize{10pt}{12pt}\selectfont
${}^{2}${\em Max Planck Institute for Physics of Complex Systems,
Dresden D-01187, Germany}}\/
}
\maketitle

\begin{abstract}
The free energy bounds for inhomogeneous current states in doped
antiferromagnetic insulators and spatial configurations of spin and  charge degrees of freedom
in lightly and heavily underdoped phases of this strongly correlated electron system are considered.
It is shown that states are characterized by a small parameter of geometric origin,
which determines the degree of packing in the knots of
filament manyfolds of
the order parameter distributions.
We found a hierarchy of the spatial scales for flat knots, which
gives rise to a free energy gain.
A toroid moment in an underdoped state is found for the first time.
It is shown that in the percolation
picture of charge density distributions, that surround
antiferromagnetic nanoclusters, the optimal level of
the hole density is in qualitative agreement with the
experimentally observed value. \\[5pt]
\end{abstract}

\section*{1. Introduction}

Doping of planar antiferromagnetic insulators breaks the antiferromagnetic order
and gives rise to the state that precedes the high-temperature
superconducting phase.
After one of ten electrons is moved, roughly speaking, to the reservoir of dopants,
the phase with short-wave length correlations of the electron degrees of freedom arises.
This strongly correlated electron system has several key properties.
Among them are strikingly small values of hole doping in the underdoped region of the phase diagram,
where there are both  unusual so-called pseudogap and superconducting phases.

The "normal" state is characterized by a
nonuniform distribution of strongly correlated spin \cite{L,L1} and charge
\cite{H,P,B,C,McEHL,HLK} degrees of freedom, as well as by an electron-hole asymmetry \cite{HLQ,H1}.
A comparison of experimental data obtained from diffent types of superconductors
shows that the universal two-component dynamics takes place
in all cuprates investigated so far. The universal multi-component
response is clear indication of the presence of the phase stratification \cite{GS} in cuprates.
Study of cooperative behavior in inhomogeneous systems is recognized nowadays
as one of the most challenging problems of the condensed matter physics \cite{KKK,KKM} and
non-Abelian field theory \cite{FN2,FN,C2,C1}.
The aim of this paper is to find bounds on the free energy of
inhomogeneous current states of doped antiferromagnetic insulators
at a moderate level of doping.

Many recent experiments provided significant support to the idea \cite{EK} that pseudogap phase
is the initial state for the superconducting phase due to precursor pairing and that it has
basically the same origin as superconductivity.
In this paper we focus  on the universal description of both phases, considering them
on equal footing. Starting from pseudogap phase, we use an equvalent form of the
spinor Ginzburg-Landau functional to study a free energy gain.
To characterize the properties of the spatial configurations of strongly correlated
spin and charge degrees of freedom in the
partial coherent phase we introduce a toroid moment \cite{DTos,M,DT}.
We show how to embed considered planar phenomena into three-dimensional space.
Using quasi-one-dimensional character of spatial distributions of hole density
and the percolation picture of the charge configurations,
we argue also why is the optimal doping in high-temperature superconductors so small.

\section*{2. Model}

To describe spatial distributions of the electron degrees of
freedom, we shall use the following $(3+0)D$ extended version \cite{BFN}
of the $O(3)$ nonlinear $\sigma$ model:
\begin{eqnarray}
  \lefteqn{F =F_{\bf n}+F_\rho+F_{\bf c}=} \\
  & & \int d^{3}x \left[\left(\frac{1}{4}\rho^{2}\left(\partial_{k}{\bf n}\right)^{2}+H_{ik}^2\right)+
      \left(\left(\partial_{k}\rho \right)^{2}-b\rho^2+\frac{d}{2}\rho^4\right) +
      \left(\frac{1}{16}\rho^{2}{\bf c}^{2}+F_{ik}^2-2F_{ik}H_{ik}\right)\right]\,. \nonumber
\end{eqnarray}
In this equation, the field of the unit vector ${\bf n}(x,y,z)$ describes
spin distributions, while the
momentum field ${\bf c}={\bf J}/\rho^{2}={\bf a}-{\bf A}$, where
${\bf J}$ is the total current, contains paramagnetic $({\bf a})$
and diamagnetic $(-{\bf A})$ parts. In Eq. (1) $F_{ik} =
\partial_{i}c_{k} -
\partial_{k}c_{i}$ and $H_{ik} = {\bf n}\cdot
[\partial_{i}{\bf n}\times \partial_{k}{\bf n}] =
\partial_{i}a_{k} - \partial_{k}a_{i}$. For simplicity,
the Higgs potential in Eq. (1) is expressed only via two constants $b$ and $d$.
The density of the current energy $\frac{1}{16}\,\rho^2{\bf c}^2$,
the surface energy $F_{ik}^{2}$
and the diamagnetic coupling $ - 2F_{ik}H_{ik}$ of the fields ${\bf a}$  and  ${\bf c}$
determine contributions to the current part, $F_{\bf c}$, of the free energy.
The dimensionless units of all the variables are given in Ref. \cite{IP}.

The first term in Eq. (1) is the energy of a Heisenberg antiferromagnet
written in the long-wave length limit.
The self-energy $H_{ik}^{2}$ of the internal magnetic
field $H_i=1/2\,\varepsilon_{iks}H_{ks}$ in the long-wave length limit
\cite{VD} includes the mean value $<0|{\bf S_1[{\bf S_2\times{\bf S_3}}]}|0>$  of
the spin noncollinearity degree on the three sites of
a plaquette of the square lattice
which appears after the
electron is moved from the fourth site to the reservoir of the dopants.
These chiral three-spin correlations of spin
degrees of freedom are the main building block for the formation of more
complicated spin and charge structures considered below. The structures are
encoded in Eq. (1) by the term $(H_{ik}-F_{ik})^{2}$ which describes also mutual
frustration of the currents ${\bf a}$ and  ${\bf c}$.

For better understanding of the role of each 
term in free energy (1), it is useful to compare 
Eq. (1) with the Hamiltonian of the $t-J$ model 
$$ 
H = \sum_{<ij> } J_{ij}\left [{\bf S}_{i}{\bf S}_{j} - \frac{1}{4}n_{i}n_{j}\right ] 
- \sum_{<ij>, \sigma } t_{ij}\left (a^{+}_{i\sigma}a_{j\sigma} + H. c.\right ) 
$$ 
in the inhomogeneous case. 
Here $a_{i\sigma} = c_{i\sigma}(1 - n_{i,-\sigma})$, 
$n_{i,\sigma} = c^{+}_{i\sigma}c_{i\sigma}$, 
${\bf S}_{i} = c^{+}_{i\alpha}{\bf \sigma_{\alpha \beta}}c_{i\beta}/2$. 
This model is a standart one when studying different phase 
states in doped antiferromagnetic insulators. At gradient expansion, 
the first and the second terms in this equation provide the 
first and the third terms in Eq. (1) respectively. 
The last kinetic term in the $t-J$ model ensures 
the existence of the term $(1/16)\rho^{2}{\bf c}^{2}$ in Eq. (1). 
The self-energy ${\bf B}^{2}/8\pi \sim {\rm curl}^{2}\,{\bf A} \sim (H_{ik} - F_{ik})^{2}$ 
of the gauge field ${\bf A}$ in Eq. (1) can be also obtained \cite{WZ,Wi} 
in the framework of the $t-J$ model. 

The exchange integral modulation, the scalar  field
$\rho (x,y,z)$ in (1) describes  the spin stiffness $\rho^{2}$ and depends by the doping level.
Indeed, the spin stiffness $\rho^2$ relates to the density $\rho_h^2$ of positively charged
vacant sites, so-called "holes", by the constraint $\rho^2+\rho_h^2=1$ \cite{IW}.
If  $\rho^{2} \lesssim 1$ , the hole density $\rho_{h}^2$, like the term $H_{ik}^{2}$, is localized
at the boundaries of the regions of the nonzero spin density, i.e. it exists on the
complement space with respect to the space of the spin density definition.
Based on this remark and on the lattice patterns of the distributions of
the spin degrees of freedom, a low-dimensional analog of the holographic
principle, which leads to the conclusion that the distribution of the
states with broken antiferromagnetic order has the boundary character,
was proposed in Ref. \cite{IP1}.

\section*{3. Knots and links of order parameters}

The properties of different phases in model (1) depend on the distributions
of competing order parameters $\bf n$, $\rho $ and $\bf c$ in the ground states.
At ${\bf c}=0$ and $\rho = const$ the configurations of the field ${\bf n}(x,y,z)$
are classified by the Hopf invariant \cite{FN1,RPKKTV,MM}
$Q={1\over16\pi^2}\int\limits_\mathcal{M} d^3x\,\varepsilon_{ikl}\,a_i\,
\partial_k\,a_l \,$.
At the compactification of  $\Bbb R^{3}\to \mathcal{M} = S^{3}$ and ${\bf n}\in S^{2}$,
an integer $Q\in Z\Leftarrow\pi_3(s^2)$ indicates the degree of linking or knotting of a
filament maniford collection in the three-dimensional space $\mathcal{M}$, where the field of the unit
vector ${\bf n}(x,y,z)$ is constant: ${\bf n}(x,y,z)={\bf n}$.
For two linked loops $Q = 1$, for the trefoil knot $Q = 6$ and etc.
In this case the local minima of the free energy $F$ are determined \cite{VK} by the inequality:
\begin{eqnarray}\label{s2}
F\ge 32\pi^2\,{|Q|}^{3/4}.
\end{eqnarray}

The quasi-one-dimensional character of the distribution of the ${\bf n}$-field
in Eq. (1) at $\rho = const$, ${\bf c}=0$, if $\rho\ne{\rm const}$, $\bf c=0$ affects also the
hole density $\rho_h^2=1-\rho^2$ \cite{IP1}. The configurations of field  $\rho_h$ in the form
of stripes and loops are most advantageous in this case.

When the field of the momentum ${\bf c}\ne 0$ is unquenched and $\rho= {\rm const}$,
free energy bound decreases \cite{PV}:
\begin{eqnarray}\label{s3}
F\ge 32\pi^2\,{|Q|}^{3/4}{\left(1-{|L|\over|Q|}\right)}^{\!2},
\end{eqnarray}
where the mutual linking number of the fields $\bf a$ and $\bf c$ equals
$L = {1\over16\pi^2}\,\int d^3x\,\varepsilon_{ksl}\,c_k\,\partial_s\,a_l \,$,
is determined by the invariant $L\not\in\Bbb Z$. The universal gain of the free energy
in Eq. (3) describes the condensation energy \cite{SKA} in the considered model of strongly
developed spin distributions.

The reduction of the free energy bound occurs due to the diamagnetic coupling
of the currents in Eq. (1). The contribution of this last term
is so significant that it consumes completely the energy $F_{\bf c}$ and a part of the energy
$F_{\bf n}$. This phase with large pair momentum differs from the FFLO state \cite{FF,LO} by
the unquenched paramagnetic part of the total current. The second distiction corresponds in a general case
to inhomogenious distribution of the total momentum of pairs existing in the incoherent
form \cite{EK} above the superconducting $T_{c}$.

The existence of quasi-one-dimensional configurations with open ends and in the form of loops in
the framework of the present model, when the free energy density has the form $f=\left(\partial_k\rho\right)^{2}+V_{eff}(\rho)+\left(F_{ik} - H_{ik}\right)^2$, with
$V_{eff} = -b_{eff}\rho^2 + (d/2)\,\rho^4$ and $b_{eff} = b -
(1/16){\bf c}^2 - (1/4)\left(\partial_k{\bf n}\right)^2$, corresponds to different phase states
\cite{P1,N}. In the phase, where the gauge symmetry remains unbroken, $b_{eff} < 0$ and  $\rho_{min}=0$.

It follows from the equation for the energy $V_{eff}$ that
inhomogeneous distributions of the fields exist only in the
case $b_{eff} > 0$. With the change of doping, approaching the phase
with $b_{eff} > 0$, where the gauge symmetry is broken due to Higgs effect, the fraction
of quasi-one-dimensional configurations with closed ends increases. In the phase $b_{eff} > 0$,
there exist \cite{N} quasi-onedimensional configurations of length $l$, satisfying the
relation $l\,\rho_{min}=2\pi n$. In the ground state ${\bf c}=0$, and the value $\rho_{min}$
for the amplitude of the Higgs field in the ground state is determined by the contribution
$(1/4)(\partial_{k}{\bf n})^{2}$ of the characteristic configurations of field ${\bf n}$.
In this case, distributions in the form of a ring are typical configurations of
field $\rho $ \cite{IP1}. For greater values $( l\,\rho_{min} > 2\pi )$ of the commensurability
parameter $l\,\rho_{min}$ in the ground state, there exist closed quasi-one-dimensional
distributions of field $\rho $. For sufficiently short $(0 < l\,\rho_{min} < 2\pi)$ configurations
of the field $\rho $ in the ground state, quasi-one-dimensional field distributions with open
ends may be found. Summarizing, we see that in magneto-inhomogeneous phases, we encounter an
internal hierarchical multi-scale structure, determined by the density field,
which becomes crucial for the description of these phases.

We see also that the state with a finite value of the momentum
field ${\bf c}$ is characterized by the competing inhomogeneous order parameters
${\bf n}$ and ${\bf c}$ with non-local correlations
\footnote{In $(2+1)D$ dimensional form of the dynamical description,
the non-local interaction between electrons
may be described using Chern-Simons statistical gauge fields $a_\alpha^i$,
which induce strong correlations of the phases of fermion quantum states at large distances.
The world line braiding matrix $k_{ij}$  in the Chern-Simons action
$S=\frac{k_{ij}}{4\pi}\int\limits_\mathcal{M} \,
d^{3}x\,\varepsilon_{\alpha \beta \gamma}
a_\alpha^i\partial_\beta\,a_\gamma^j$ in this case is proportional
to the matrix $K_{\alpha \beta}$ (see below), describing
the knotting degree of the field configurations of the $(3+0)D$ free energy.} 
between them due to the entanglement
of their one-dimensional distributions.
Note that this conclusion is in agreement with one of the main statements
of Refs. \cite{VGKT,VGK1} that inhomogeneous current states cannot
be described with the aid of Ginzburg-Landau functional
with the one-component order parameter. In our case the existence of inhomogeneous
current states is due to the non-linear coupling of several order parameters
${\bf n},\,{\bf c}$ and $\rho$ at the level of their equations of motion.
The analogous conclusion may be made, starting from a
lattice description of current states in a so-called staggered flux
phase \cite{LNNW}, considering strong correlations of charge and spin
degrees of freedom in the mean field approach. 
Flux phases and $DDV$ states \cite{HR,Sh,Na}, i.e. the phases with a modulation 
of the current density, are well-known particular examples of the toroid states. 
In this paper we focus our attention  on
non-perturbative analysis, and consider a general case when all
inhomogeneous order parameters are present.

\section*{4. Inhomogeneous current states}.

The basic property of model (1) is the compactness of field  ${\bf n}$,
which leads to its one-dimensional distributions in the form
of knots \cite{FN1,RPKKTV,MM,GH,BS,HS}, and to the discrete character
of spectrum (2).
To analyze the general case of ${\bf c}\not=0$ and $\rho \not=const$,
we first write down  the energy $F_{\bf n}$ at $\rho=const$ and $x_i\to x_i\,r$
in the form of $F_{\bf n}=\int\,d^{3}x\,\left[g_1\left(\partial_k{\bf n}
\right)^2+g_2\,H_{ik}^{2}\right]=\sqrt{g_{1}g_{2}}\left(r/R+R/r\right)$, where $r$
is the scale factor, $R=\sqrt{g_2/g_1}$ and $g_{1} \sim \rho ^{2}$ and $g_2 =1 $ are the coupling constants.
The latter equation shows that the characteristic knot size at the minimum
of the energy  $F_{\bf n}$ is proportional to $\rho^{-1}$.

With the increase of doping $\rho_h^2$, i. e. with
decrease of $\rho^2$, the radius $R\sim 1/\rho$ grows up to its certain critical value
$R_{cr}$. Beginning from this radius, the configurations of field
${\bf n}$, forming the knot, are unstable \cite{W} with respect to the scale
transformations.
This means that knots of a size greater than
$R_{cr}$ collapse, and form localized structures near the points of the stereographic
projection $\Bbb R^3\to S^3$. This phenomenon appears to be the key one, when
we compare different contributions to the free energy (1) in an inhomogeneous
current state at small enough values of $\rho^2$.

We see that at sufficiently small density $\rho^2$ or (which is the same)
at sufficiently high level of doping $\rho_h^2$,
beginning from the finite values of density $\rho^{2}$, in order
to describe strongly correlated charge and spin degrees of freedom,
one may introduce a small parameter, determined by the relation $R_0/R$,
where $R_0$ is the correlation length of the ${\bf n}$- and ${\bf c}$-field distributions.
The radius $R_0$ has the geometric meaning of the filament thickness, for which
the configurations of these fields are determined. A typical value
of $R_0$ is of the order of 10 \AA, 
i. e. it equals three-four lattice constants.
Since Ginzburg parameter in this case is much less than unity, we may
use the mean field approach (1). A characteristic scale $R$ has
the meaning \cite{GGKV} of the effective radius of current correlations.

Let us assume that $R_0 < R < R_{cr}$ and compare the decrease of the free energy
(1) due to the diamagnetic term $ - 2F_{ik}H_{ik}$ with its increase
due to the term  $(\partial_k\rho )^2$.
If the contribution of the diamagnetic term $-2F_{ik}H_{ik}$
is of the order of $c_0/R_{0}^3$,
where $c_0$ is the amplitude of the momentum field ${\bf c}$,
the estimate of the maximum contribution of the term $(\partial_k\rho)^2$
yields $1/(RR_0)^2$.
The gain obviously should be also greater than the contributions of other
terms in the last bracket in Eq. (1), i. e. of the surface term
$F_{ik}^2\sim c_0^2/R_0^2$ as well as of the current energy
$\rho^2{\bf c}^2\sim F_{ik}^2(R_0/R)^2$.
Combining these inequalities, we see that at $\alpha=(R_0/R)^2 << 1$ and at a finite
value of the momentum determined by the inequality
 \begin{equation}
  \alpha/R_0 < c_0 < 1/R_0
 \end{equation}
even in a state with the inhomogeneous distribution of the density  $\rho^2$
the free energy may be smaller due to diamagnetism than at ${\bf c}=0$ in Eq. (1).
The parameter $\alpha $ has the meaning of the degree of filament packing in the
three-dimensional space. It is the {\it searched small parameter} in our problem of strongly
correlated spin and charge degrees of freedom when all coupling constants are of the order
of unity. In the case ${\bf n} = const$,
it is proportional to $\kappa^{-2}$ where $\kappa $ is the Ginzburg-Landau parameter.

To find the condition for a state with momentum from the range (4) when
the energy is less, than at ${\bf c}=0$ (in spite of inhomogeneity), let us compare the
gain of the total free energy with the contribution from
the term $(\partial_k\rho)^2$.
The loss of the energy is of the order of $R_z/R^2$.
Here  $R_z$ is the "vertical" knot size.
Therefore, the diamagnetic contribution is the main one for sufficiently
flat knots \cite{MHDKK}, existing at $R_0 < R_z < R <R_{cr}$.
This means that the energy of configurations in the sector with the
topological invariant $L \not=0$ even in the inhomogeneous state may be less
than at ${\bf c}=0$.
This conclusion is one of the main results of the paper.
The condensation energy  \cite{SKA}, which depends on the
linking degrees $L$ and $Q$, decreases in comparison
with the case of $\rho = const$ \cite{PV}. The minimal value of the free energy
gain $-\Delta F$ can be estimated as $-\Delta F \simeq \left(64\pi^{2}/Q^{1/4}\right)\alpha$.

\section*{4. Toroid state}

Let us assume that the level of doping is small enough, i.e. \ $\rho_h \ll 1$.
Then the spin density $\rho^2$ is large and a knot size $R$ is so small that
$\alpha \lesssim 1$. The decrease of the degree $\alpha$  of filament packing in the knot
with the increase of $\rho_h$ is accompanied by the transition to a state, in which
among planar field projections the field configurations $\rho(x,y)$ in the form of closed
one-dimensional distributions
are most preferable \cite{IP1}.

The fact, whether the boundary charge states at the boundaries of these
isulating regions will form an infinite percolative cluster, depends on the hole density $\rho_h^{2}$.
Let the level of doping be insufficient for the appearance of an extended connectable cluster.
In accordance with Ginzburg's proposal, we will call this phase state with the spontaneous
diamagnetism a toroid state \cite{VGK1}.
The phase state, in which current order is characterized by a polar vector $T$,
that changes sign under time reversal,
can be represented by the ordering of toroid moments.
It is characterized by a toroid moment ${\bf T}$,
which is defined in the following way \cite{DTos,M}
\begin{equation}\label{s4}
\bf a={\rm curl}\,\,{\rm curl}\,\,\bf T \, .
\end{equation}
Using the equality (4) and the identity \cite{MH}
\begin{equation}\label{s5}
{({\rm curl}\,{\bf a})}_i={1\over2}\,\varepsilon_{ikl}\,{\bf n}
\left[\partial_k{\bf n}\times\partial_l{\bf n}\right]
\end{equation}
one can express the toroid moment via the chiral degree of noncollinearity
$\displaystyle{\bf n}\cdot\left[\partial_k{\bf n}\times\partial_l{\bf n}\right]$ in the form:
\begin{equation}\label{s6}
T_i({\bf r})={1\over16\pi^2}\int d^3r'\left[
{({\bf r}-{\bf r}^{\,\prime})_k\over{|{\bf r}-
{\bf r}^{\,\prime}|}^3}\int d^3r''\,
{{\bf n}\cdot\left[\partial_k{\bf n}\times\partial_i{\bf n}\right]
\over|{\bf r}^{\,\prime}-{\bf r}^{\,\prime\prime}|}\right].
\end{equation}

We see that toroid moment ${\bf T}$ is given by the non-local Biot-Savart law as well as by the
factor which is determined by the Coulomb Green function. The vector ${\bf T}$ characterizes the
distribution of the poloidal component of the current on the torus \cite{DTos,M}.
The magnetic
field of this current is confined to the interior of the torus. For the degree of noncollinearity
(localized at a point) $\displaystyle {\bf n}\cdot\left[\partial_k{\bf n}
\times\partial_i{\bf n}\right]=\varepsilon_{ki}\,\delta({\bf r})$ of the spin distributions,
the toroid moment has the form
\begin{equation}\label{s7}
T_i({\bf r})={\varepsilon_{ki}\over16\pi^2}\,\int
d^3r'\;{{({\bf r}-{\bf r}^{\,\prime})}_k\over
r'\,{|{\bf r}-{\bf r}^{\,\prime}|}^3}=
{\varepsilon_{ki}\,x_k\over 8\pi\,r} \, .
\end{equation}
If the chiral density $\displaystyle {\bf n}\cdot\left[\partial_k{\bf n}
\times\partial_i{\bf n}\right]=\left(\varepsilon_{ki}/(2\pi R)\right)\,\delta (z)\delta (r-R)$
of spin distributions is localized in the plane $z=0$ on a ring
of radius $R$, the toroidal moment equals $T_{i}(r,z=0)=\varepsilon_{ki}(x_{i}/r)t(r)$ with
the function $t(r)$ having its
maximum at $r=R$ and $t(r) \sim r^{-2}$ at $r \gg R$.

In the toroid state the diamagnetic susceptibility $\chi^{\prime}$
is determined \cite{VGK1} by the radius $R$ of current correlations in such a way that
$\chi^{\prime} = \chi_{L}/\alpha$. Here $\chi_L=-e^{2}k_{F}/(12\pi ^{2}mc^{2})$ is
the Landau diamagnetic susceptibility for non-interacting electrons.
For porous knots with $\alpha \ll 1$, it can be rather significant.
Anomalously great diamagnetic susceptibility is caused by a great radius $R$ of current correlations.
It should also be noted that the condition \cite{GGKV} $\chi_{L} \sim -e^{2}nr^{2}/(mc^{2}) > -1/4\pi $ 
equivalent to the inequality $r < r_{D} =c/\omega_{p}$ 
once again attract our attention to the boundary character 
of the considered phenomena. Here $n$ is the mean electron density, $r_{D}$ is the Debye radius, 
and $\omega_{p} = (4\pi ne^{2}/m)^{1/2}$ is the plasma frequency. 
We pay attention to the fact that in this phase the total magnetic
moment of clusters equals zero and the toroid state is $d$-{\it symmetric}.

The realization of the toroid state is achieved in systems belonging to 31 classes of
magnetic symmetry where the vector ${\bf T}$ can exist \cite{GGKV}.
The conditions \cite{GGKV} for the realization
of a toroid state are easier to satisfy (i) in the neihborhood of inhomogeneities and
(ii) in systems with strong electron correlations. (iii) The toroidal state formed against the
background of the antiferromagnetic state is the most likely situation
as well as (iv) conditions of the presence of charged and magnetic impurities,
i.e. holes and dynamical non-collinear spin distributions in our case. (v) Strong spin-orbit interaction
also gives rise to effects associated with toroidal ordering.
All these conditions are met in the system at hand.

\section*{5. Planar phenomena in $3D$ space}

Under the conditions when all the coupling constants in (1) are of the order of unity
\footnote{This is one of the reasons, why the problem discussed cannot be currently solved, even numerically.}
the main object of the study appears to be the geometric properties and the
topology of knotted one-dimensional manifolds
of the order parameter domain, which give the main contribution to
the free energy. One of the results of this approach is the
appearance of the parameter $\alpha=(R_0/R)^2<1$, characterizing the degree of filament
packing in the knot. The existence of the minimal value of the field
$\rho ^{2}$, playing the role of the running coupling constant in Eq. (1), i. e.
the existence of the minimal value
$\alpha_{min}=(R_{0}/R_{cr})^{2}$, is similar to the emergence of the
solution for the superconducting gap in a two-component model
\cite{LW}, beginning (unlike BCS model) with the finite value of the coupling constant.

Let us describe a method of embedding planar phenomena into the $3D$ space.
We choose a tuning parameter $\rho^2\sim 1/R^2$
that $\alpha_{min} < \alpha << 1$. Let us consider the projection
of complementary spaces for knots which are pulled over the knots
(so-called Seifert surfaces) to the plane  $(x,y)$.
For definiteness, we consider the simplest (among torus knots) trefoil knot
with $Q=6$.
In the case of the flat  $\,$ ($R_z < R$) knot,
the complementary space is a triple twisted band.
In the case of the knot stretched along the axis  $z$ (at $R < R_z < R_{cr}$)
the Seifert surface is a torus with a hole (see Fig. 1)
\footnote{The braided filaments of the trefoil, stretched along the time
axis, one can consider in $(2+1)D$ dynamical case as the result of the
action of $b_1^3$ on the quantum states of the braid group operator $b_1$.
}.

\begin{figure}
\begin{center}
\epsfxsize=100 mm \epsfbox{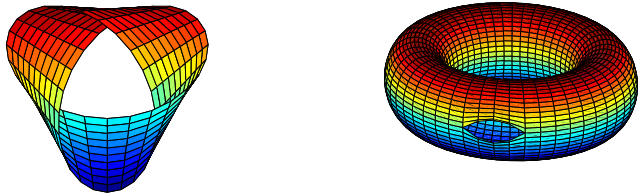}
\end{center}
\caption{(colour online). Complementary surfaces for a flat (l.h.s.) and extended along $z$-axis (r.h.s.)
trefoil knots.
}
\end{figure}

The charge degrees of freedom are distributed (as it has been already mentioned)
at the boundary of the complementary space for knots.
Therefore, the boundary line presents a circle, which is the projection of these
one-dimensional boundary configurations for a sufficiently flat knot.
For the stretched knot (at $R < R_z < R_{cr}$), the line is a stripe with the length of the order
of $R_0$, equal to the scale of the hole mentioned above.
Summarizing, we can see that the free energy gain in an inhomogeneous
current state is achieved, due to diamagnetism, for the distributions
of the charge density on the plane in the form of loops.

The study of such configurations of the charge density is based on 
measuring the density of states (DOS) by means of scanning tunneling microscope (STM).
In such experiments \cite{KHKATT} the formation of the DOS filaments in the form of rings
have been discovered recently in the entangled labyrinthine pattern
of quasi-one-dimensional distributions of the charge density in
planar compounds $Ca_{2-x}Na_{x}CuO_{2}Cl_2$ at the moderate
doping $x$.

\section*{6. Optimal doping}

Whether the phase both
with $\alpha_{cr} < \alpha << 1$ and the finite pair momentum
\cite{FF,LO} ${\bf c}_0$ remains superconducting or appears to be
an intermediate state with spontaneous diamagnetism \cite{GGKV}
depends perhaps on the connectness of the total current cluster
(see e.g. \cite{G}), formed by the projections of Seifert surfaces
on the plane.

It is natural to suppose that the occurrence of pre-formed electron pairs in the underdoped state \cite{EK}
and a phase coherence in the superconducting phase are indepent phenomena.
Starting with the picture of pre-formed pairs, we discuss in this section some
features of the phase coherence due to percolation \cite{MKM} using charge density structures mentioned above.
With the increase of the hole density $\rho _{h}^{2}$, the size of ring-shaped current distributions increases
to such an extent that the conditions arise for the appearance of two-dimensional connectable current
clusters and the transition to the state with ${\bf c} \neq 0$ with the gain of free energy.
We shall find the level $p_{best}$, for which the temperature of the transition to the superconducting state
is maximal.

If we did not take into account the distributions of the charge degrees of freedom at the
boundaries of the filament configurations of the field ${\bf n}$ in the knot only, we would deal
with $2D$ percolation problem with a well-known \cite{G} answer $p_{cr}=0.5$ for the bond percolation
threshold value. A considerable distinction between the experimentally observed value $p_{best,\,exp}=0.16$
from this value is a convenient test for checking the accepted point
of view.

To solve the problem in question we shall use the recent experimental results \cite{KHKATT}.
With the aid of the scanning tunnel microscope it was found in \cite{KHKATT} that
among quasi-one-dimensional configurations with an increased density of charge states
in the underdoped region, the fraction of filament field distributions in the
form of ring-shaped structures increases with the increase of the doping level. Using the
corresponding trial functions of paper \cite{IP1} it was shown that in the framework of model (1),
such configurations of the field $\rho _{h}$ are preferable when the degree of doping increases.
The interpretation of the results of measuring the distributions of the spin degrees of freedom
in underdoped states on the basis of neutron scattering \cite{L,L1} is also based on the assumption
of quasi-one-dimensional nanoscale distributions in the planes of conducting clusters in the form
of thin rings.

Let us suppose that the boundary currents are distributed along the boundaries of ring insulator
regions with such a relationship that the ring thickness differs from its diameter, as it follows
from the experiment \cite{KHKATT}, by one order $(\alpha^{1/2}\simeq 0.1)$. The geometrical confinement
allows us to neglect the contribution of internal (with respect to ring-shaped) distributions of degrees of freedom. Thus for the threshold of the flow along the connectable cluster,
formed by ideal distributions in the form of rings, we get a considerabl smaller value:
$p_{best} = 0.5-0.314 = 0.186$ in comparison with the standart one $p_{cr}=0.5$. For a current
cell in the form of a hexahedron, $p_{best}=0.179$. This example shows that comparatively slight
deviations of the form of the elementary current cell from the ideal ring-shaped form may lead to
a shift in the value $p_{best}$ towards the experimentally observed value $p_{best,\,exp}=0.16$.
If a square, constructed by the distributions in the form of stripes, were used
instead of a ring, we would have $p_{best}=0.5-0.4=0.1$.

Study of general properties of high-temperature superconducting states, belonging to different classes
of compaunds, shows that the dependence of the critical temperature on the hole concentration, for example
in $Y_{1-x}Ca_{x}Ba_{2}Cu_{3}O_{6}$ and $La_{2-x}Sr_{x}Cu)_{4}$,
is well represented \cite{TBSHJ,T} by the empiric curve
$T_{c}/T_{c,max} = 1-82.6(p-0.16)^{2}$ at $p=x/2$. This parabolic dependence $T_{c}(p)$, has zeros at
$p_{min} \approx 0.05$,  $p_{max} \approx 0.27$ and a maximum at $p_{best}=0.16$. Optimally doped
compounds based on thallium, for example $Tl_{0.5}Pb_{0.5}Sr_{2}Ca_{0.8}Y_{0.2}Cu_{2}O_{7}$, have
$p_{best}=0.15$. Some other classes of high-temperature superconductors point to similar values.
We note also that the underdoped pseudo-gap phase with strong developed spin and charge distributions
appears in the above mentioned classes of the compounds at a well-defined \cite{B1}
hole concentration $p \approx 0.19$.

\section*{6. Conclusion}

Under conditions of moderate doping $\rho_h^2$, the
value of $\rho^2$ may be such that the knot size $R \sim 1/\rho\sim R_z$
appears to be comparable with the correlation length  $R_0$.
Since all values in Eq. (1) are of the same
order, these greater values of $\rho^2$ correspond to
smaller values of the momentum ${\bf c}$ and due to Eq. (3) the energy gain is less.
In the optimum case of great values
${\bf c}$ the self-dual state $F_{\bf n} \approx F_{\bf c}$,
which is characterized by the dense packing of knot filaments
($\alpha \lesssim 1$), ${\bf a}-{\bf c} = {\bf A} = 0$ and by the matrix \cite{AK}
 $K_{\alpha\beta} =
\frac{1}{16\pi^{2}}\int\limits_\mathcal{M} d^{3}x \,
\varepsilon_{ikl}a_{i}^{\alpha}\partial_{k}a_{l}^{\beta} =
 \left( \begin{array}{cc}
 Q & L  \\
 L & Q^{\prime}
 \end{array} \right) \Rightarrow
 Q \left( \begin{array}{cc}
 1 & 1  \\
 1 & 1
 \end{array} \right)$, where $a_i^1 \equiv a_i$, $a_i^2 \equiv c_i$,
may be considered as initial compressible state for the transition
to the incompressible phase $\alpha << 1$ with the increase of the
doping degree. The  compressibility of the state at $\alpha
\lesssim 1$ occurs due to the existence of non-zero Poisson
bracket $\{x_{0},y_{0}\}$ for the coordinates $x_{0},y_{0}$ of the
vortex filament center. The uncertainty of the location of the
core center at the dense filament packing gives rise to the
nullification of the linking degree  $Q$ due to the effect of the
filament reconnection. This leads to the transition to the ground
state with $Q=0$.

The comparison of the results of the
present paper with the conclusions of \cite{S}, following from the
application of the conformal field theory methods to find the
exact solution of the problem of superconducting states in cluster
systems, requires thorough analysis of the details of the
nullification ($k+2 = 0$) mechanism of the linking coefficients $k+2 \sim Q$ in
$SU(2)$ Chern-Simons action.

In conclusion, we have compared the value of the free energy gain
due to diamagnetism in the inhomogeneous current state with
the value of the energy loss due to inhomogeneity and found the conditions,
under which the energy of inhomogeneous states with a current appears
to be less than in the absence of the current. Studying the threshold level
of doping with the use of the percolation approach of describing current configurations,
existing in doped antiferromagnetic insulators, we have found, that in the toroid
state the optimal value of the hole concentration is close to the experimentally observed
value. The toroid moment in this $d$-symmetric state is expressed via the
chiral degree of noncollinearity of the distribution of the spin degrees of freedom. 

\section*{Acknowledgments}

We are grateful to L. N. Bulaevskii, Yu. M. Gal'perin, L. D. Faddeev, P. Fulde, Yu. V. Kopaev, 
R. Ramazashvili, V. A. Rubakov, S. R. Shenoy and H. Takagi for helpful discussions and advices. 
This work was supported in part by the RFBR under Grant No. 04-02-16684. \\\\

\end{document}